\documentclass[onecolumn,showpacs]{revtex4}
\usepackage{graphicx}
\usepackage{dcolumn}
\usepackage{bm}
\usepackage{dcolumn}\DeclareRobustCommand\bblash{\btt{\@backslashchar}}
\usepackage{amsmath}\makeatother
\usepackage{epsfig}

\begin{document}

\title{Dynamics of Quintessential Inflation}

\author{Zhai Xiang-hua }
 \email{kychz@shnu.edu.cn}
\author{Zhao Yi-bin }%

\affiliation{%
Shanghai United Center for Astrophysics(SUCA), Shanghai Normal
University, 100 Guilin Road, Shanghai 200234,China\\
}%
\date{\today}

\begin{abstract}
In this paper, we study a realistic model of quintessential
inflation with radiation and matter. By the analysis of the
dynamical system and numerical work about the evolution of the
equation of state and cosmic density parameter, we show that this
model is a good match for the current astronomical observation.
The conclusion we obtain is in favor of the model where the
modular part of complex field plays the role of the inflaton
whereas the argument part is the quintessence field. The numerical
calculation shows that a heteroclinic orbit (solution of dynamical
system) interpolates between early-time de Sitter phase (an
unstable critical point) and a late-time de Sitter attractor.\\

\end{abstract}

 \maketitle

\section{Introduction}
Astronomical observation on the Cosmic Microwave Background(CMB)
Anisotropy \cite{CMB}, Supernova Type Ia(SNIa) \cite{SNIa} and
SLOAN Digital Sky Survey(SDSS) \cite{SDSS} converge on that our
Universe is spatially flat, with about seventy percent of the
total density resulting from dark energy that has an equation of
state $w<-1/3$ and drives the accelerating expansion of the
Universe which began at a redshift of order one-half. The origin
of the dark energy remains elusive from the point of view of
general relativity and standard particle physics. Several
candidates to represent dark energy have been suggested and
confronted with observation: cosmological constant, quintessence
with a single field \cite{SingleField} or with N coupled field
\cite{NCoupledField}, phantom field with canonical \cite{Caldwell}
or Born-Infield type Lagrangian \cite{BIL}, k-essence
\cite{k-essence} and generalized Chaplygin gas(GCG) \cite{GCG}.
Among these models, the most typical ones are cosmological
constant and quintessence which has caught much attention ever
since its invention.

The idea of inflation is legitimately regarded as an great
advancement of modern cosmology: it solves the horizon, flatness
and monopole problem, and it provides a mechanism for the
generation of density perturbations needed to seed the formation
of structures in the universe \cite{Kolb&Turner}. In standard
inflationary models \cite{Liddle&Lyth}, the physics lies in the
inflation potential. The underlying dynamics is simply that of a
single scalar field rolling in its potential. This scenario is
generically referred to as chaotic inflation in reference to its
choice of initial conditions. This picture is widely favored
because of its simplicity and has received by far the most
attention to date. The properties of inflationary models are also
tightly constraint by the recent result from the observation. The
standard inflationary $\Lambda$CDM model provides a good fit to
the observed cosmic microwave background (CMB) anisotropy. Peeples
and Vilenkin proposed and quantitatively analyzed the intriguing
idea \cite{QuintInf} that a substantial fraction of the present
cosmic energy density could reside in the vacuum potential energy
of the scalar field responsible for inflation (quintessential
inflation). After that, there were some models to be presented in
succession. However, most of these models involve
non-renormalizable, special potentials and these potentials
usually do not have a local minimum, making the conventional
reheating process unoperative.

On the other hand, the Peccei-Quinn (PQ) symmetry
\cite{Peccei&Quinn} is the most elegant solution to the strong CP
problem of QCD. The global $U(1)_{PQ}$ is a spontaneously symmetry
breaking (SSB). Weinberg \cite{Weinberg} and Wilczek
\cite{Wilczek} pointed out that there is a Nambu-Goldstone boson,
\textquoteleft\textquoteleft the
axion\textquoteright\textquoteright, associated with SSB. Such a
global symmetry often arises in supersymmetric \cite{Gu&Li} and
superstring-inspired models \cite{Li&Gu}. A priori PQ symmetry
breaking scale is arbitrary, which can be taken a value between
$10^2$ Gev and $10^{19}$ Gev. In most axion models PQ symmetry
breaking happens at a temperature $T \sim f$, when a complex
scalar field $\Phi$, which carries $U(1)_{PQ}$ charge, develops a
vacuum expectation value.  At temperatures below $T \sim f$ the
renormalizable potential for $\Phi$ is

\begin{equation}
V_1(\eta) = \frac{\lambda}{4} (\eta^2 - f^2)^2,
\end{equation}
where $\eta = (\Phi^\ast \Phi)^{\frac{1}{2}}$. However, the
argument of $\langle \Phi \rangle$ is left undecided when the
$U(1)_{PQ}$ has been broken because of $V_1(\eta)$ is independent
of $\textbf{arg}\langle \Phi \rangle$. The massless $\xi$ degree
of freedom is the axion: $\xi =
(f/\widetilde{N})\textbf{arg}\langle \Phi \rangle$, where
$\widetilde{N}$ is the color anomaly of the PQ symmetry
\cite{Kaplan}. The axion is massless at $T \gg {QCD} \; scale $,
but the axion develops a mass due to instanton effects at low
temperatures. In the $\widetilde{N} = 1$ case, the axion mass is
easy to concretize by using a potential

\begin{equation}
V_2(\xi) = M^4 [1- \cos(\xi/f)],
\end{equation}
where $M^4 = m_\xi^2 f^2$ and the axion develops a mass $m_\xi$.
The conception that the modular part of a complex field with PQ
symmetry may drive inflation is also not new \cite{PiBabuBarr}.
Another version of a complex field was proposed as a realistic
candidate for inflation \cite{NatInf} and, separately, for
quintessence \cite{Quint} in a natural fashion. The two-axion
model has also been suggested for quintessence \cite{quintaxion}
or inflation \cite{twoaxion}. The axion model can also be used for
quintessence \cite{Gauge} and, separately, for inflation
\cite{Extra} from extra dimensions.

 Recently, a fascinating model for
quintessential inflation has been proposed by Rosenfeld and Frieman
(RF) \cite{Rosenfeld&Frieman} with aid of the axion theory. In their
model, a renormalizable complex scalar field $\Phi$ described by a
lagrangian with a $U(1)$ symmetry spontaneously broken at a high
energy scale $ f \sim 10^{19}$ Gev and explicitly broken by
instanton effects at a much lower energy can account for both the
early inflationary phase and the recent accelerated expansion of the
Universe. The modular and argument parts of the field $\Phi$ were
identified with the inflaton and the quintessence fields
respectively.

In this paper, we put the emphasis on the study of the dynamics of
RF model with radiation and matter composition. According to the
phase space analysis and numerical calculation, we show that how
the modular part of complex field plays the role of the inflaton
whereas the argument part is the quintessence field. Phase space
methods are particularly useful when the equations of motion are
hard to solve analytically for the presence of radiation and dust
matter. In fact, the numerical solutions with random initial
conditions are not a satisfying alternative because of these may
not reveal all the important properties. Therefore, combining the
information from the critical points analysis with numerical
solutions, one is able to give the complete classification of
solutions according to their early-time and late-time behavior.
There are stable and unstable critical points for RF model with
radiation and dust matter composition. Critical points are always
exact constant solution in the context of autonomous dynamical
systems. These points are often the extreme points of the orbits
and therefore describe the asymptotic behavior. In our case, the
unstable critical points are corresponding to a de Sitter behavior
at very first moments of the Universe. The stable critical points
are corresponding to the second stage of accelerated expansion
began at a redshift $z \sim 1.5$ and is still operative, which
alleviates the fine tuning problem. The equation of state $w$
varies with the cosmic evolution and approaches towards -1
asymptotically showing the existence of a cosmological constant at
late times. If the solutions interpolate between critical points
they can be divided into heteroclinic orbit and homoclinic orbit.
In our case, the heteroclinic orbit connects unstable and stable
critical points. The numerical calculation shows that brought the
Universe back to the usual Friedman-Robertson-Walker expansion,
then the second  stage of accelerated expansion began at $z \sim
1.5$.

\maketitle
\section{Phase Space of $U(1)$ quintessential inflation}
For convenience, we investigate firstly the global structure of
the dynamical system without radiation and matter. The equations
of motion for RF model \cite{Rosenfeld&Frieman} are
\begin{align}\label{eq:motion}
\begin{split}
&\ddot{\eta} + 3 H \dot{\eta} - \frac{\dot{\xi}^2}{f^2} \eta +
V_1'(\eta) = 0,\\
&\ddot{\xi} + \left(3 H + 2\frac{\dot{\eta}}{\eta} \right)
\dot{\xi} + \frac{f^2}{\eta^2} V_2'(\xi) = 0,
\end{split}
\end{align}
and the Einstein equation is

\begin{equation}\label{eq:einstein}
H^2=\frac{\kappa^2}{3}(\rho_\eta +\rho_\xi ),
\end{equation}

\noindent where $H$ equals ${\dot a}/{a}$ is Hubble parameter and
$\kappa^2=8\pi G$. Here we define the energy densities:
\begin{align}\label{definition}
\begin{split}
&\rho_\eta=\frac{1}{2}{\dot
\eta}^2+V_1(\eta),\\
&\rho_\xi=\frac{1}{2}\frac{\eta^2}{f^2}{\dot \xi}^2+V_2(\xi).
\end{split}
\end{align}

To gain more insights into the dynamical system, we introduce the
new variables
\begin{align}\label{newvar}
\begin{split}
u&=\eta, \quad x=\xi,\\
v&=\dot\eta, \quad y=\dot{\xi},
\end{split}
\end{align}
then the equations of motion could be reduced to
\begin{align} \label{eq:auto1}
\begin{split}
\frac{du}{dt}&=v,\\
\frac{dv}{dt}&=-3Hv+\frac{u y^2}{f^2}-V_1'(u),\\
\frac{dx}{dt}&=y,\\
\frac{dy}{dt}&=-(3H+\frac{2 v}{u})y-\frac{f^2}{u^2} V_2'(x).
\end{split}
\end{align}
And $H$ could be rewritten as

\begin{equation}
H=\Bigl[\frac{\kappa^2}{3}\left(\frac{1}{2}v^2+V_1(u)+\frac{1}{2}\frac{u^2}{f^2}y^2
+V_2(x)\right) \Bigr]^{1/2}
\end{equation}
Here we have $V_1(u)=\frac{\lambda}{4}(u^2-f^2)^2$ and
$V_2(x)=M^4[1-\cos(x/f)]$.

Before we carry out the numerical study, we would like to analyze
the system of equations qualitatively. The critical point of the
autonomous system Eqs.(\ref{eq:auto1}) is $(u_c,0,x_c,0)$, where
$u_c,x_c$ are defined by $V_1'(u_c)=0, V_2'(x_c)=0$ respectively.
So one can obtain the critical points:
\begin{equation}\label{critpoint}
(u_c,v_c,x_c,y_c)=(\pm f,0,k f \pi,0),
\end{equation}
where $k$ is any integer.

In order to investigate the stability of these critical points, we
can linearize the Eqs.(\ref{eq:auto1}) about the critical points

\begin{align}
\dot{\bar{y}}&=\textbf{A} \bar{y}, \\
\textbf{A}&=\frac{\partial f}{\partial \bar{y}}(u_c,v_c,x_c,y_c).
\end{align}
Here $\bar{y}=(u,v,x,y)^T$ and $f$ is short for the right hand
sides of Eqs.(\ref{eq:auto1}). Then we have the stability matrix
$\textbf{A}$

\begin{equation}
\begin{pmatrix}
0  &   1   &   0   &   0  \\
-V_1''(u_c) &  -\sqrt{3 \kappa^2 V_2(x_c) } &  0  &  0 \\
0  &   0   &   0   &   1  \\
0  &   0   &   -V_2''(x_c) & -\sqrt{3 \kappa^2 V_2(x_c)}
\end{pmatrix}.\nonumber
\end{equation}
The eigenvalues of the stability matrix are

\begin{align}\label{eigenvalue}
\begin{split}
\Bigl[\; &\frac{1}{2}(-\sqrt{3 \kappa^2 V_2(x_c)}-\sqrt{3 \kappa^2
V_2(x_c)-4V_1''(u_c)},\\
&\frac{1}{2}(-\sqrt{3 \kappa^2 V_2(x_c)}+\sqrt{3 \kappa^2
V_2(x_c)-4V_1''(u_c)},\\
&\frac{1}{2}(-\sqrt{3 \kappa^2 V_2(x_c)}-\sqrt{3 \kappa^2
V_2(x_c)-4V_2''(x_c)},\\
&\frac{1}{2}(-\sqrt{3 \kappa^2 V_2(x_c)}+\sqrt{3 \kappa^2
V_2(x_c)-4V_2''(x_c)} \; \Bigr].
\end{split}
\end{align}
It is clear that the critical points are always stable when $k$ is
even because the real part of all the eigenvalues is negative. So
these critical points correspond to late-time attractor solutions.
When $k$ is odd, the fourth eigenvalue in Eq.(\ref{eigenvalue}) is
always larger than zero and the corresponding critical points are
unstable, which are exact constant solution with de Sitter behavior
at early-time of the Universe. This is to say that the dynamical
system has stable critical points at the minimum of the potential
$V_1(\eta)$ and the minimum of the potential $V_2(\xi)$. Next, let's
read out the physical implications when the system is at the
critical point regime. The cosmic parameters are

\begin{equation}\label{omega}
\Omega_\eta=\frac{\rho_\eta}{\rho_c} , \quad
\Omega_\xi=\frac{\rho_\xi}{\rho_c},
\end{equation}
and the equation of state of the complex field is
\begin{equation}\label{w}
w=\frac{v^2+\frac{u^2}{f^2}
y^2-2V_1(u)-2V_2(x)}{2(\rho_\eta+\rho_\xi)}.
\end{equation}
Clearly, from Eqs.(\ref{omega}) and (\ref{w}), one can find that
$w=-1$, $\Omega_\eta=0$ and $\Omega_\xi=1$ at the late time
attractor.

\begin{figure}[h]
\epsfig{file=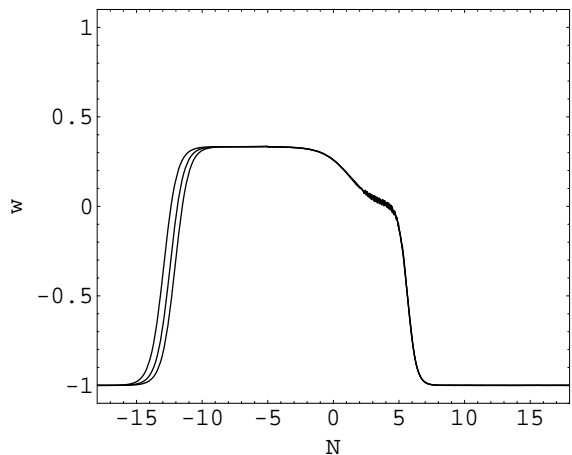,height=2.5in,width=3in} \caption{Evolution of
the equation of state at the presence of radiation and matter.The
curves correspond to the different initial conditions.}\label{fig1}
\end{figure}

\begin{figure}[h]
\epsfig{file=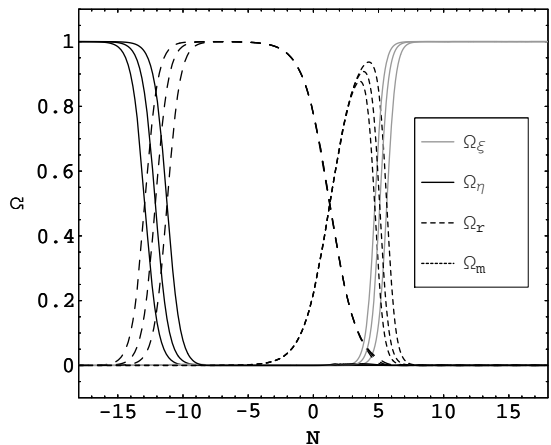,height=2.5in,width=3in} \caption{Evolution of
cosmic density parameter of quintessence energy density
$\Omega_\xi$, inflaton energy density $\Omega_\eta$, radiation
energy density $\Omega_r$ and matter energy density $\Omega_m$.The
curves correspond to the different initial conditions}\label{fig2}
\end{figure}

\section{Evolution at the presence of radiation and matter}
If we are to search for a realistic model, we must know exactly how
the complex scalar field $\Phi$ couples to ordinary matter. For
convenience, we assume that coupling is very weak following
traditional ideas. In fact, the most significant feature of all the
axion couplings is that they are proportional to $\widetilde{N}/f$:
the larger the PQ SSB scale, the weaker the axion couples.
Furthermore, we have taken the $f \sim 10^{19}$ Gev so that couples
can be neglected. Certainly, the ordinary matter still affect the
evolution of $\Phi$ via their contribution to the general expansion
of the Universe.

Now we solve the equations of motion via a numerical calculation
at the presence of radiation and matter and obtain the results
that will confirm our qualitative analysis. To do this, we
introduce the new variable
\begin{equation}
N=\ln a.
\end{equation}
Therefore, the Eqs.(\ref{eq:auto1}) are reduced to
\begin{align} \label{eq:auto2}
\begin{split}
\frac{du}{dN}&=\frac{v}{H},\\
\frac{dv}{dN}&=-3v+\frac{u y^2}{f^2 H}-\frac{V_1'(u)}{H},\\
\frac{dx}{dN}&=\frac{y}{H},\\
\frac{dy}{dN}&=-(3+\frac{2 v}{u H})y-\frac{f^2 V_2'(x)}{u^2 H}.
\end{split}
\end{align}
and
\begin{equation}\label{H}
H^2=H_i^2E^2(N),
\end{equation}

\noindent where $H_i$ denote the Hubble parameter at an initial
time. $E(N)$ is defined \cite{Hao&Li} as
\begin{equation}\label{en}
\begin{split}
E(N)=\Bigl[\frac{\kappa^2}{3H_i^2}\left(\frac{1}{2}v^2+V_1(u)+\frac{1}{2}\frac{u^2}{f^2}y^2
+V_2(x)\right)\\
+\Omega_{r, i}e^{-4N}+\Omega_{m,i}e^{-3N}\Bigr]^{1/2}.
\end{split}
\end{equation}
$\Omega_{r, i}$ and $\Omega_{m, i}$ are the cosmic density
parameters for  radiation and dust matter at the initial time. Note
that when N goes to be very large, or at late-time, the contribution
to $H$ from matter and radiation will become negligible.

 Solving these equations will give us some insights into the
evolution of the fields and the quantities of interest. The
corresponding conclusions of numerical calculation are shown in
Fig.\ref{fig1} and Fig.\ref{fig2}. This numerical solution describes
a heteroclinic orbit, which interpolate between an unstable de
Sitter critical point and a late-time de Sitter attractor.
Therefore, this model is more natural to explain the two stages of
acceleration. A point worth emphasizing is that the ordinary matter
(radiation and dust) affect the evolution of the scalar field via
their contribution to the general expansion of the Universe because
of the couples can be neglected in the model. In Fig.\ref{fig1}, the
behavior of the equation of state parameter is shown. At initial
time, $w=-1$ corresponds the de Sitter expansion (inflationary
phase), then it increases and becomes positive. After arriving at
the value $1/3$, the Universe comes to the radiation dominated epoch
and $w$ stays on a broad platform. Next, $w$ drops to zero and stays
on a narrow platform. Finally, $w$ drops below zero and approaches
to $-1$, which corresponds second stage of accelerated expansion.
The evolution of cosmic density parameters are shown in
Fig.\ref{fig2}. The argument part contribution $\Omega_\xi$ stays at
$\Omega=0$ at the very first moments of the Universe and becomes $1$
in late-time, which plays the role of quintessence field. On the
other hand, the modular part contribution $\Omega_\eta$ plays the
role of inflaton. During the whole evolution, the radiation energy
density and the dust matter energy density become dominate
respectively. Therefore, we see that the constraints arising from
cosmological nucleosynthesis and structure formation are satisfied.

\section{Conclusion and discussion}

In this paper, we have discussed the cosmological implication of a
complex scalar model of quintessence inflation with radiation and
matter. Using the numerical calculation for this model, we show that
the argument contribution is negligible at the early epoch of the
universe while it becomes dominate with the time evolving and the
evolution of the modular part is the opposite, which is a viable way
to unify the description of the inflationary stage and the current
accelerated expansion. Analysis to the dynamical evolution of the
complex scalar model indicates that it admits a late-time attractor
solution, at which the field behaves as a cosmological constant.
Obviously, attractor and heteroclinic orbit are both alleviate the
fine tuning problem. We can also extend our discussion to the
lagrangian with non-abelian symmetry, which is an interesting
extended work and we will discuss in another preparing work.

\section*{Acknowledgments}
This work is supported by National Natural Science Foundation of
China under Grant No. 10473007 and  by Shanghai Rising-Star
Program under Grant No. 02QA14033.

\end{document}